\documentclass[twoside,11pt]{article}
\usepackage{cite,amsfonts,amssymb,exscale,bbm,a4wide}
\newcommand{\cl}[1]{\mbox{${\cal #1}$}}

\def\emph#1{{\sl #1\/}}
\def\R{{\mathbbm R}}
\def\Z{{\mathbbm Z}}

\newcommand {\be}{\begin{equation}}
\newcommand {\e}{\end{equation}}
\newcommand {\bea}{\begin{eqnarray}}
\newcommand {\ea}{\end{eqnarray}}
\newcommand {\nit}{\noindent}

\newcommand {\bit}{\bibitem}

\newcommand {\alga}{{\mathfrak a}}

\newcommand {\algd}{{\mathfrak d}}
\newcommand {\alge}{{\mathfrak e}}
\newcommand {\f}{{\mathfrak f}}
\newcommand {\g}{{\mathfrak g}}

\newcommand {\da}{\delta}

\newcommand {\fract}[2]{\mbox{${\textstyle{\frac{#1}{#2}}}$}}

\def\ie{{\sl i.e.\/}}
\def\eg{{\sl e.g.\/}}

\def\dopreprint{\hfill{\small\thepreprint}\\}%
\def\preprint#1{\def\thepreprint{#1}}%
\def\thepreprint#1{}%
\def\address#1{\date{{\sl #1}\\\ \\\theversion}\gdef\date##1{}}%
\def\version#1{\gdef\theversion{#1}}%

\version{16 December 2002}%
\preprint{DAMTP-2002-161}

\begin{document}
%

\title{\dopreprint Development of a unified tensor calculus for the\\
                   exceptional Lie algebras}
\author{A.~J.~Macfarlane${}^1$\thanks{e-mail:
A.J.Macfarlane@damtp.cam.ac.uk}\ \ and
        Hendryk Pfeiffer${}^{2,3,1}$\thanks{e-mail:
H.Pfeiffer@damtp.cam.ac.uk}}
\address{\small ${}^1$ Centre for Mathematical Sciences, DAMTP,
Wilberforce Road, Cambridge CB3 0WA, UK\\
         ${}^2$ Perimeter Institute for Theoretical Physics, 35 King
Street N, Waterloo ON N2J 2W9, Canada\\
         ${}^3$ Emmanuel College, St.~Andrew's Street, Cambridge CB2
3AP, UK}

\vspace{1cm}
\date{\version}

\maketitle

%
%

\begin{abstract}

The uniformity of the decomposition law, for a family $\cl{F}$ of Lie
algebras which includes the exceptional Lie algebras, of the tensor
powers $ad^{\otimes n}$ of their adjoint representations $ad$ is now
well-known. This paper uses it to embark on the development of a
unified tensor calculus for the exceptional Lie algebras. It deals
explicitly with all the tensors that arise at the $n=2$ stage,
obtaining a large body of systematic information about their
properties and identities satisfied by them. Some results at the $n=3$
level are obtained, including a simple derivation of the the dimension
and Casimir eigenvalue data for all the constituents of $ad^{\otimes
3}$. This is vital input data for treating the set of all tensors
that enter the picture at the $n=3$ level, following a path already
known to be viable for $\alga_1 \in \cl{F}$.

The special way in which the Lie algebra $\algd_4$ conforms to its place
in the family $\cl{F}$ alongside the exceptional Lie algebras is
described.

\end{abstract}

%
\section{Introduction}
%

\subsection{Notation and conventions}

Let $\g$ be a simple complex Lie algebra with generators $X_k$ such that
\be \label{a1}
{[} X_i \, , \, X_j {]} =i c_{ijk} X_k . \e \nit
Here and in the following summation over repeated indices is
understood. The adjoint representation $ad$ of $\g$ is defined by
$X_k\mapsto ad_k:=[X_k,\,\cdot\,]$ with matrix elements
\be \label{a2} (ad_i)_{jk} = ic_{ikj} , \e \nit
and our normalisations are fixed by requiring that the Cartan-Killing
form of $\g$ satisfies
\be \label{a3}
\kappa_{jk}= {\rm Tr} \; (ad_j ad_k) =\da_{jk} , \e \nit
so that the structure constants are totally antisymmetric and satisfy
\be \label{a4}
c_{jpq} \; c_{kpq} =\da_{jk} . \e \nit
It follows that the quadratic Casimir operator of $\g$
\be \label{a5}
{\cal C}^{(2)}=X_k \; X_k , \e \nit
has, for each $\g$, the eigenvalue $c_2(ad)=1$, since
\be \label{a6}
{\cal C}^{(2)}(ad)_{ij}= c_{pqi} \; c_{pqj} =\da_{ij} .\e \nit

We use the notation $\cl{E}$ to indicate the set of all exceptional Lie
algebras
\be \label {a7} \cl{E}= \{ \g_2, \f_4, \alge_6, \alge_7, \alge_8 \}. \e
\nit
They are a subset of the set
\be \label{a8} \cl{F}=\{ \alga_1, \alga_2, \g_2, \algd_4, \f_4, \alge_6,
\alge_7, \alge_8 \} \e \nit
of Lie algebras in the last line of the extended Freudenthal magic
square \cite{freu,ff}.

We note that we use the informal abbreviation irrep for irreducible
representation. Here the term is
understood as irreducible over the field of
complex numbers, but only up to diagram automorphisms. The groups of
diagram automorphisms of the algebras $\g\in\cl{F}$ are $\Z_2$ for
$\alga_2$ and $\alge_6$, $S_3$ for $\algd_4$, and the trivial group
for all the others. As the adjoint irrep is always mapped to itself
under diagram automorphisms, we find in the complete decomposition of
its tensor products over the complex numbers either irreps that are
self-conjugate or pairs of conjugate irreps for $\alga_2$ and
$\alge_6$. For $\algd_4$, the constituents are either single irreps
that are stable under triality or triples and sextuples of irreps that
are related by triality. We call the direct sum of all irreps
belonging to such a pair, triple or sextuple, an \emph{irrep in the
sense of this article}.

We refer informally to a ''Clebsch" as an abbreviation for a
Clebsch-Gordan coefficient. We thought the abbreviations were
marginally preferable to the acronyms IR and CGC. We refer to irreps
usually by their dimensions because our studies give a central role to
dimension formulas for families of irreps, one for each of $\g \in
\cl{F}$, as a function of $D=\dim \; \g$. When we wish to use the
Dynkin co-ordinate or highest weight designation of an irrep, we
follow the conventions that stem from the choice of Cartan matrix used
in \cite{fss} and \cite{cor}.

\subsection{On the context and the content of this paper}

The work of Meyberg\cite{meyAH} for $j=2$, and its extension
\cite{del,CdM} to $j=3$ and $4$, demonstrates the uniformity for $\g
\in \cl{F}$ of the decomposition into irreps of the $j$-th tensor
power
\be \label{a21} ad^{\otimes j} \e \nit
of the adjoint irrep $ad$ of $\g \in \cl{F}$. This striking property
opens the way towards the
main purpose of the present paper: the development of a comprehensive
tensor
calculus for the exceptional Lie algebras $\cl{E}$,
in a form uniform over  $\cl{E}$.
The first phase of this programme is implemented here, mainly but not
exclusively, at the $j=2$ level.

We have in an earlier paper \cite{ajmg2} succeeded in developing quite
far
a tensor calculus for $\g_2$, one that is based, not on $ad$ as here,
but on
the seven dimensional defining irrep of $\g_2$. This can no doubt serve
satisfactorily applications to areas like spin-chains, Gaudin models
or to integrable quantal or supersymmetric models with $\g_2$
invariance.
However it promises no  discernible path to a comparable treatment of
the larger exceptional groups, nor is it as amenable, as are the methods
of the
present paper, to extension beyond the context of the
two-fold direct product. The uniform tensor calculus that this paper
displays underlines the view that the tensor powers (\ref{a21}) of $ad$
for
$\g \in \cl{E}$ have very special properties that deserve to be,
and here will be, exploited as fully as possible.

The complete reduction of
\be \label{I2}
ad \otimes ad = (ad \otimes ad)_A + (ad \otimes ad)_S , \e \nit
provides a result for the antisymmetric part
\be \label{I3}
(ad \otimes ad)_A \equiv ad + X_2 , \e \nit
in a universal form, one that is valid for each simple complex $\g$,
and a  result
\be \label{I4}
(ad \otimes ad)_S \equiv R_1 + R_2 + R_3 , \e \nit
valid for $\g \in \cl{F}$ but not for other $\g$.
It is a simple matter to show that $X_2$ is an irrep of $\g$ in the
sense of this article with the universal
properties
\bea  \Delta_2 \equiv \dim \; X_2 & = & \fract{1}{2} D(D-3) \nonumber \\
       c_2(X_2) & = & 2,  \label{I5} \ea \nit
where $D=\dim\g$ and $c_2(R)$ denotes the eigenvalue of the quadratic
Casimir operator
$\cl{C}^{(2)}$ for the irrep $R$ of $\g$. (we use the notations $X_2$
and $X_3$
below, see (\ref{h9}), following \cite{del}, but sometimes write
$X_2=R_0$ and
$X_3=R_7$ to enable generic reference to representations $R_r$.)
The membership for $\g \in \cl{F}$ of the family of irreps $X_2$  is
shown in
Table~\ref{tab_x2}, along with the membership of the three families
$R_1, R_2$ and $R_3$ arising in (\ref{I4}).

\begin{table}
\begin{center}
\begin{tabular}{|c|c|c|c|c|c|}
\hline
           & $\g_2$ & $\f_4$ & $\alge_6$
              & $\alge_7$ & $\alge_8$ \\ \hline \hline
    $ad$   &  14   &  52  &  78   & 133 & 248  \\
           & (10) & (1000) & $(000001)$ & $(1000000)$ & $(00000010)$ \\
    $c_2$  &  1     &  1 &  1     & 1 & 1    \\ \hline
    $X_2$  &  $77^\prime$   &  1274 &  2925  & 8645 & 30380  \\
           & (03) & (0100) & $(001000)$ & $(0100000)$ & $(00000100)$ \\
    $c_2$  &  2     &  2  &  2     & 2 & 2  \\ \hline
    $R_1$  &  1    &  1     &  1     & 1        & 1     \\
    $c_2$  &   0     &  0      &  0     & 0     & 0  \\ \hline
    $R_2$  &  27    &  324    &  650    & 1539 &  3875  \\
           & (02) & (0002) & $(100010)$ & $(0000100)$ & $(10000000)$ \\
    $c_2$  &   \fract{7}{6} &  \fract{13}{9} &  \fract{3}{2} &
    \fract{14}{9}  &  \fract{8}{5}    \\ \hline
    $R_3$  & 77 & 1053  & 2430 & 7371   &  27000  \\
           & (20) & (2000) & $(000002)$ & $(2000000)$ & $(00000020)$  \\
    $c_2$  & \fract{5}{2} & \fract{20}{9} &
    \fract{13}{6}  &  \fract{19}{9} & \fract{31}{15} \\ \hline
\end{tabular}
\end{center}
\caption{\label{tab_x2} Irreps of $\g$ for $ad\otimes ad$.}
\end{table}

It is known \cite{meyAH} that there are formulas for
\be \label{I6} D_2= \dim \; R_2, \quad  D_3= \dim \; R_3,
\quad c_2(R_2) =\fract{1}{2}(1+\ell_2), \quad c_2(R_3)
=\fract{1}{2}(1+\ell_3),
\e \nit
as functions of $D=\dim \; \g$, valid in each case for each member of
the
family in question. Eq. (\ref{I6}) also defines the useful variables
$\ell_{2,3}$;
$D_{2,3}$ and $\ell_{2,3}$ are essential as input into many important
formulas derived below.

While we have favoured the use of $D$ as a parameter in formulas valid
across
a family of irreps of $\cl{F}$, other parameters are in common use
elsewhere.
To facilitate comparisons, we have collected some information about them
in the Appendix.

We review the analysis of $ad \otimes ad$ carefully in
Sec.~\ref{sect_level2}, which makes various additions to results in
\cite{meyAH}, and employs an elementary method that lends itself to
generalisation beyond the case of $ad \otimes ad$.  One matter of
interest that arises here is the absence \cite{ok2} for $\g \in
\cl{E}$ of a primitive quartic Casimir operator. We review this too,
giving some extra results of later use, and explain how the case of
$\algd_4$, which has two independent primitive quartic Casimir
operators, nevertheless conforms fully to the family picture for
$\cl{F}$. Apart from this topic we are concerned almost exclusively
with the exceptionals.

The remaining sections of this paper take up the establishment, first
at the $j=2$ level, of a tensor calculus applicable uniformly across
$\cl{E}$, and the demonstration that our methods are sufficiently
general as to permit extension to the case of $ad \otimes ad \otimes
ad$, although a systematic treatment of this is left to a future
publication.

In Sec.~\ref{sect_simple}, we embark on the tensor calculus accessible
using tensor products such as $v_i v_j$ and $\phi_i \phi_j$, where
$v_i$ is an adjoint vector, \ie\ one that transforms according
to the adjoint representation of $\g \in \cl{E}$, and $\phi_i$ is an
adjoint vector with anticommuting components, \eg\ fermionic
creation operators. This brings into focus many of the most important
third rank isotropic tensors, these being isotropic under adjoint
action. We deduce basic identities involving them, and consider the
interpretation of them as Clebsches. Further, we note the occurrence
of results like those which in the quantum theory of angular momentum
see the appearance of Racah coefficients, or what is the same thing to
within a phase of no significance here, Wigner $6j$ symbols.

Just as for full mastery of $\alga_n=su(n+1)$ tensor and related
algebraic methods,
stems from development of the properties both of the Gell-mann
$\lambda$-matrices and the tensors that enter their product law
\be \label{a31}
\lambda_i \lambda_j= \frac{2}{n} \delta_{ij}+(d+if)_{ijk} \lambda_k , \e
\nit
so also is there a best approach to the tensor calculus for $\g \in
\cl{E}$.

Thus in Sec.~\ref{sect_matrices}, we introduce a basis of matrices for
$\hom_\R (\cl{V}_{ad},\cl{V}_{ad})$, where $\cl{V}_{ad}$ is the vector
space in which $ad$ acts.  It is not surprising that we meet a much
more complicated situation when we attempt to generalise from
$\alga_n$ to the set of exceptionals $\cl{E}$ treated uniformly. We
write and confront fully the set of all product laws involving
matrices of the basis, thereby identifying all the isotropic third
rank tensors arising at the $j=2$ level of our study. Their
interpretation as Clebsches is discussed. Amongst the large body of
identities, applicable uniformly across $\cl{E}$, that are proved in
Sec 4.  are again some that have an interpretation in terms of Racah
coefficients.  Further we identify explicitly (and evaluate quadratic
Casimir operators for) the matrices that transform under $\g \in
\cl{E}$ according to the irreps $R_2, R_3$ and $X_2$. This is
essential input into Sec.~\ref{sect_level3}.

In Sec.~\ref{sect_level3}, we study $ad \otimes ad \otimes ad$,
following a straightforward method for deducing, for all $g \in
\cl{E}$, formulas for the dimensions and the quadratic Casimir
eigenvalues for all the (new) families of irreps of $g \in \cl{E}$
that enter $ad \otimes R$ for $R=X_2, R_2, R_3$.

This is vital input data for a study of all the tensors that enter the
picture for $ad^{\otimes 3}$. It is hoped to describe progress in this
direction in a future publication. That a viable approach is available
is known from a preliminary study for the simplest case of $\alga_1
\in \cl{F}$, but details of this are not given here.

In Sec.~\ref{sect_d4}, we make some remarks regarding the status of
$\algd_4$ within $\cl{F}$. We show explicitly how the fact that
$\algd_4$ has two primitive quartic Casimir operators, whereas all
other members of $\cl{F}$ have none, is fully compatible with this
status.

The paper concludes with three appendices. The first describes the
parametrisation of family formulas by $m$ instead of $D=\dim \; \g$,
the second gives a listing of dimension formulas in terms of $m$,
whilst the third compares the definition of Racah coefficients, or
$6j$ symbols, in quantum theory of angular momentum with various
formulas derived in the text expressing products of three trilinear
tensors in terms of one. The relevance of this follows from a view
described in Sec.~\ref{sect_matrices} of such tensors as
Clebsch-Gordan coefficients.

%
\section{Analysis of $ad \otimes ad$}
%
\label{sect_level2}

\subsection{The $L$-operator}
\label{sect_l}

\begin{table}
\begin{center}
\begin{tabular}{|c|c|c|c|c|c|c|} \hline
   ${}$ & $\alga_2$ & $\g_2$ & $\f_4$ & $\alge_6$
              & $\alge_7$ & $\alge_8$\\ \hline \hline
     $\ell_2$ & $-\fract{1}{2}$ & $- \fract{5}{12}$    &
$-\fract{5}{18}$
&  $-\fract{1}{4}$     & $-\fract{2}{9}$  &  $-\fract{1}{5}$   \\ \hline
   $\ell_3$  & \fract{1}{3} & \fract{1}{4} & \fract{1}{9}   &
\fract{1}{12}  &
\fract{1}{18}    &    \fract{1}{30}    \\ \hline
\end{tabular}
\end{center}
\caption{\label{tab_l} Eigenvalues of $L$}
\end{table}

Our approach here employs the $L$-operator for $\g$. Writing
$X_{1i}=X_i\otimes I$ and $X_{2i}=I\otimes X_2$, this is defined, for
$X_i= X_{1i}+X_{2i}$ , with $X_{1i} \mapsto ad_{1i}, X_{2i} \mapsto
ad_{2i}$, by writing
\be \label{c12}
{\cal C}^{(2)}(ad \otimes ad)=(X_{1i}+X_{2i})(X_{1i}+X_{2i})
={\cal C}^{(2)}(ad)+2L+{\cal C}^{(2)}(ad) \; \e \nit
in agreement with (\ref{a5}). Clearly $L=ad_{1i} ad_{2i}$
has the same
eigenspaces as ${\cal C}^{(2)}(ad \otimes ad)$
so that its eigenvalues are given by
\be \label{c13}
\ell=\fract{1}{2} c_2(ad \otimes ad) -1 .\e \nit
We have $\ell=-1,
-\fract{1}{2}, 0$ for $R_1, ad, X_2$ in all cases, and for $R_2$ and
$R_3$
we have the eigenvalues given in Table~\ref{tab_l}. From these data one
can see empirically the result
\be \label{c14}
\ell_2 + \ell_3 + \fract{1}{6}=0, \e
which we derive in Sec.~\ref{sect_matrices}.

\subsection{Trace results}

From (\ref{a2}) and (\ref{c12}), we get
\be \label{x1}  L_{ij,pq}=-c_{ipk} c_{jqk}. \e
Thus we have the following trace results
\bea {\rm Tr} \; I & = & D^2 \nonumber \\
{\rm Tr} \; L & = & 0 \nonumber \\
{\rm Tr} \; L^2 & = & D \nonumber \\
{\rm Tr} \; L^3 & = & -\fract{1}{4}D \label{x2} \ea
The third result here comes from (\ref{a6}), while the fourth one
depends on
the consequence
\be \label{x3} c_{ipj} c_{jqk} c_{kri} = - \fract{1}{2} c_{pqr} \e
of the Jacobi identity. Such a result is valid for all $\g$, but the
actual
number on the right side depends on our conventions, (\ref{a4}) and
(\ref{a6}).

Alongside (\ref{x3}) we note the result
\be \label{x4}
c_{ipj} c_{jqk} c_{kri} c_{pqs} =  - \fract{1}{2} \delta_{rs}. \e \nit

\subsection{Projectors}

We begin by stating a well-known result. If a hermitian operator $A$ has
distinct
eigenvalues $a_i$, $1\leq i\leq p$, then the projector onto its $i$-th
eigenspace is given by
\be \label{d2}
P_i= \prod_{k \neq i} \frac{A-a_kI}{a_i-a_k}  \e \nit
where $I$ is the unit operator. Further the result
\be \label{d3}
(A-a_iI) P_i=0  , \e \nit
with no sum implied on $i$, is, for each $i$, a possibly reduced version
of the
characteristic equation of $A$. Rather than employ the full $L$-operator
of
(\ref{c12}), with unit operator $I$ such that
\be \label{d4}
(I)_{ij,pq}= \da_{ip} \da_{jq} , \e \nit
we treat separately the symmetric and antisymmetric parts of $L$
\be \label{d5}
L_S= LI_S \quad , \quad L_A=LI_A , \e \nit
the corresponding unit operators being
\be \label{d6}
(I_S)_{ij,pq}=\fract{1}{2}(\da_{ip} \da_{jq}+\da_{iq} \da_{jp}) , \quad
(I_A)_{ij,pq}=\fract{1}{2}(\da_{ip} \da_{jq}-\da_{iq} \da_{jp}) . \e
\nit
Then in notation corresponding to (\ref{I3}) and (\ref{I4}), we have
\be \label{d7}
P_{ad}+P_0= I_A \quad , \quad P_1+P_2+P_3=I_S. \e \nit
We turn now first to $L_A$ and second to $L_S$.

\subsection{The antisymmetric subspace}

We wish to note the universal features of the results for $L_A=LI_A$.
The result $\ell_{ad}=-\fract{1}{2}$ for $ad$ is obvious for any
$\g$. From the Jacobi identity (see, for example, \cite{MaPf00}) we
obtain,
\begin{equation}
\label{e4}
  L_A=-\fract{1}{2} P_{ad},
\end{equation}
and also the reduced characteristic equation of $L_A$,
\begin{equation}
\label{e5}
  L_A(L_A+\fract{1}{2})=0,
\end{equation}
which can be rewritten as
\begin{equation}
\label{e6}
  L(L+\fract{1}{2})I_A=0.
\end{equation}
This implies that $L$ has got two distinct eigenvalues $-1/2$ and $0$
on the antisymmetric subspace, therefore $\ell_0=0$.

Although much of the most important information about $L$ for our
purposes
resides in $L_S$, (\ref{e6}) helps us simplify our work.

We note one other result. Let $\Delta_2={\rm dim}\; X_2$, then
(\ref{I3})
leads to
\be \label{e7}
{\rm Tr}\; I_A= \fract{1}{2} D(D-1) =D+\Delta_2 , \e \nit
so that
\be \label{e8}
\Delta_2=\fract{1}{2} D(D-3) , \e \nit
holds for all simple $\g$. For $\alga_2= su(3)$ this means
$\Delta_2=20$, where $20 \equiv 10 + {\overline{10}}$ is irreducible
in the sense of our paper, namely a pair of conjugate irreps.

Since easily
\be  \label{e9} {\rm Tr}\; LI_A=-\fract{1}{2} D , \e
equation (\ref{e6}) yields
\be \label{e10} {\rm Tr}\; L^2I_A=\fract{1}{4} D, \quad
{\rm Tr}\; L^3I_A=-\fract{1}{8} D. \e
This and (\ref{x2}) provide all the trace results for $L$
needed  below.

\subsection{The symmetric subspace}

We recall the formula in (\ref{x3}) for $L$, and the result (\ref{d6})
for
$I_S$. We know trivially
\be \label{f1}
(P_1)_{ij,pq}=\fract{1}{D} \da_{ij} \da_{pq}  . \e \nit
To derive, for each of the exceptional Lie algebras, which are  governed
by
(\ref{I4}), an identity quartic in the structure constants, we must
eliminate
$P_2$ and $P_3$ from the equations
\be \label{f3}
(L- \ell_i) P_i=0, \quad \ell_1=-1 \quad , \quad i=1,2,3 . \e \nit
There is no sum on $i$ in (\ref{f3}). Also $L$ can therein be replaced
by
$LI_S$ because $I_SP_i=P_i$ for each of the three symmetric projectors
$P_i$.
Since $P_2+P_3= I_S-P_1$ ,
it is easy to find the result
\be \label{f4}
(L- \ell_2)(L- \ell_3)(I_S-P_1)=0 . \e \nit
To reach the sought after identities, we use $\ell_1=-1$ and expand
(\ref{f4})
\be \label{f5}
L^2 I_S -(\ell_2+\ell_3) LI_S+\ell_2\ell_3I_S-(1+\ell_2)(1+\ell_3)P_1=0
, \e \nit
add $L^2I_A$ to this using (\ref{e6}), and obtain
\be \label{f6}
L^2= (\ell_2+\ell_3) LI_S-\fract{1}{2}
LI_A-\ell_2\ell_3I_S+(1+\ell_2)(1+\ell_3)P_1 . \e
\nit
Now the $ij,pq$ matrix element of (\ref{f6}) yields
\bea
& {} & c_{lir}c_{ljs} c_{krp} c_{ksq}= {\rm Tr}\; (ad_i ad_p ad_q ad_j)
\label{f7} \\
& {} & \qquad
=-\fract{1}{2} (\ell_2+\ell_3-\fract{1}{2})c_{lip} c_{ljq}
+\fract{1}{2} (\ell_2+\ell_3+\fract{1}{2}) c_{liq}  c_{lpj}
-\fract{1}{2} \ell_2 \ell_3 (\da_{ip} \da_{jq}+\da_{iq} \da_{jp})
\nonumber \\
& {} & \qquad +\fract{1}{D} (1+\ell_2)(1+\ell_3) \da_{ij} \da_{pq}
\nonumber . \ea \nit
We have already put $\ell_1=-1$ into this result. If we make use also of
(\ref{c14}), we may simplify the final expression on the right side of
(\ref{f7}) obtaining
\be \label{f8}
\fract{1}{3} c_{lip} c_{ljq}+\fract{1}{6} c_{liq} c_{lpj}
-\fract{1}{2} \ell_2 \ell_3 (\da_{ip} \da_{jq}+\da_{iq} \da_{jp})
+ \fract{1}{6D} \left( 5 +6\ell_2 \ell_3 \right) \da_{ij} \da_{pq}.
\e \nit
From this we may deduce the result
\be \label{f9}
{\rm Tr}\; (ad_{(i} ad_p ad_q ad_{j)}\, )=
\frac{1}{6D} [5 +6(1-D) \ell_2 \ell_3] \da_{(ij} \da_{pq)} , \e \nit
where the enclosure of a set of suffices by round brackets indicate
symmetrisation at unit weight.
This result is used below in the discussion of the non-primitivity of
the
quartic Casimir invariants of exceptional algebras.

\subsection{Derivation of formulas for $\ell_2$ and $\ell_3$}
\label{sect_ells}

The results of Sec.~\ref{sect_matrices} and~\ref{sect_level3} can be
given in a nice form, applicable uniformly to all exceptional $\g$, by
deriving formulas for $\ell_2+\ell_3$ and $\ell_2 \ell_3$ in terms of
$D={\rm dim}\;\g$.

Thus we examine the equations
\bea
{\rm Tr}\; I_S & = & \fract{1}{2} D(D+1)= 1+D_2+D_3 \nonumber \\
{\rm Tr}\; L_S & = & \fract{1}{2} D= -1+\ell_2 D_2+\ell_3 D_3 \nonumber
\\
{\rm Tr}\; L^2 I_S & = & \fract{3}{4} D= 1+{\ell_2}^2 D_2+{\ell_3}^2 D_3
\nonumber \\
{\rm Tr}\; L^3I_S & = & -\fract{1}{8} D =-1+{\ell_2}^3 D_2+{\ell_3}^3
D_3 ,
\label{g1} \ea \nit
in which $D_i={\rm dim} \; R_i$, $i=1,2,3$ with $D_1=1$. Also we have
set
$\ell_1=-1$. We write the first two and the last two of these equations
in a
matrix form
\bea
\frac{D+2}{2} \left( \begin{array} {c} D-1 \\ 1 \end{array} \right)
& = & \left( \begin{array} {cc} 1 & 1 \\ \ell_2 & \ell_3 \end{array}
\right)
\left( \begin{array} {c} D_2 \\ D_3 \end{array} \right)
\nonumber \\
\left( \begin{array} {c} (3D-4)/4 \\ (8-D)/8 \end{array} \right)
& = & \left( \begin{array} {cc} \ell_2{}^2 & \ell_3{}^2 \\
\ell_2{}^3 & \ell_3{}^3 \end{array} \right)
\left( \begin{array} {c} D_2 \\ D_3 \end{array} \right) . \label{g2} \ea
\nit Elimination of $D_2$ and $D_3$ matrixwise leads to
\bea
\frac{3D-4}{2(D+2)} & = & -x(D-1)+y \nonumber \\
\frac{8-D}{4(D+2)} & = & y[-x(D-1)+y]-x, \label{g3} \ea \nit
where $y=\ell_2+\ell_3$ and $x=\ell_2 \ell_3$,
and hence to $y=-1/6$, as noted empirically. The result
\be \label{g4}
\ell_2 \ell_3= -\frac{5}{3(D+2)} , \e \nit
follows easily, and hence expressions for $\ell_2, \ell_3, D_2, D_3$ as
explicit
functions of $D$. These are displayed in Sec.~\ref{sect_explicit}.

The results given in Table~\ref{tab_l} are all in agreement with
(\ref{g4}).
The important trace result (\ref{f7})
can now be given in a form in which
all numerical coefficients determined solely by $D$. Also (\ref{f9}) now
reads
\be \label{g40}
{\rm Tr}\; (ad_{(i} ad_p ad_q ad_{j)}\, )=
\frac{5}{2(D+2)} \da_{(ij} \da_{pq)} . \e \nit

\subsection{Explicit results}
\label{sect_explicit}

To make explicit as functions of $D=\dim \g$ some
results given in previous subsections, we define
\be \label{A1} \Delta= \big[ \frac{242+D}{2+D} \big]^{\fract{1}{2}}, \e
\nit
 denoted by $w$ in \cite{meyAH}.
This takes these values for $\g \in \cl{F}$:
\be \label{A2} 7, 5, 4, 3, \fract{7}{3}, 2, \fract{5}{3}, \fract{7}{5}.
\e \nit
Then, from Sec.~\ref{sect_ells}, we have the following
\bea \ell_2 & = & \frac{(-1-\Delta)}{12} , \nonumber \\
\ell_3 & = & \frac{(-1+\Delta)}{12} , \nonumber \\
D_2 & = & \frac{(D+2)}{4\Delta} (-D-11+\Delta(D-1)) \nonumber \\
D_3 & = & \frac{(D+2)}{4\Delta} ( D+11+\Delta(D-1)) \nonumber \\
c_2(R_2) & = & \frac{1}{6}(11-\Delta) \nonumber \\
c_2(R_3) & = & \frac{1}{6}(11+\Delta). \label{A3} \ea \nit

Since $\Delta= \frac{m+6}{m+2}$ follows (\ref{K5}) and (\ref{A1}), all
the
formulas listed here are rational functions of $m$. Thus it will often
be
true that simplifications are easier to find by working in terms of $m$
rather than $D$.

\subsection{Quartic Casimir operators for the exceptionals}
\label{sect_quartic}

We deal here with $\g \in \cl{F}$ excluding $\algd_4$ which requires
separate
treatment, provided in Sec.~\ref{sect_d4}.

The simplest thing to do in this context is to define the general
adjoint
matrix $A=b_i \; ad_i$, $b_i \in \R$, and look at ${\rm Tr}\; A^4$.
We may use (\ref{g4}) to deduce
\be \label{j1}
{\rm Tr}\; A^4=b_i b_j b_p b_q {\rm Tr} (ad_{(i} ad_j ad_p ad_{q)})\,
=\frac{5}{2(D+2)} (b_k b_k)^2 , \e \nit
which exhibits explicitly the failure of ${\rm Tr}\; A^4$ to be
primitive.

More generally, we must define a quartic Casimir operator. Set $M=ad_i
X_i$,
where the $X_i$ denote the hermitian
Lie algebra generators themselves, and define
\bea
{\cal C}^{(4)} & = & {\rm Tr}\; M^4 \nonumber \\
 & = & {\rm Tr}\; (ad_i ad_p ad_q ad_j) X_i X_p X_q X_j . \label{j2}
\ea \nit
We employ (\ref{a3}), with $\ell_2+\ell_3$ and $\ell_2\ell_3$ given by
(\ref{c14}) and
(\ref{g3}), to evaluate  the right side (\ref{j2}). We can complete
the evaluation with the aid of (\ref{a1}), (\ref{a6}) and (\ref{x3}),
obtaining
\be \label{j3} {\cal C}^{(4)}= \frac{5}{2(D+2)} {\cal C}^{(2)}{}^2
+\frac{D-3}{12(D+2)} {\cal C}^{(2)}, \e
valid for all exceptional $\g$.

\subsection{$\cl{C}^{(4)}$ for irreps of $\g \in \cl{E}$ and ${\rm Tr}\;
L^4$}
\label{sect_c4}

Let $X_i \mapsto M_i$ define the matrices of an irrep $R$ of $\g \in
\cl{E}$
with $L$-operator $L=ad_t X_t$ represented by
\bea L_M & = & ad_t M_t \nonumber \\
(L_M)_{ja,kb}  & = & (ad_t)_{jk} (M_t)_{ab} \nonumber \\
& = & -c_{tjk} m_{tab} , \label{ff1} \ea \nit
where $(M_t)_{ab} = -im_{tab}, \quad a, b, \in \{ 1, 2, \dots , \dim \;
R \}$.

The properties (\ref{a4}) and (\ref{x3}) allow us to evaluate traces
\bea {\rm Tr}\; L_M & = & 0 \nonumber \\
{\rm Tr}\; L_M{}^2 & = & m_{tab} m_{tab} = {\rm Tr}\; M_t M_t =c_2(R)
\dim \; R
\nonumber \\
{\rm Tr}\; L_M{}^3 & = & -\fract{1}{4} {\rm Tr}\; M_t M_t .\label{ff2}
\ea \nit
Also
\bea
{\rm Tr}\; L_M{}^4 & = & {\rm Tr}\; (ad_j ad_k ad_q ad_p)
{\rm Tr}\; (M_j M_k M_q M_p) \nonumber \\
& = & \left\{ \frac{5}{(D+2)} c_2(R)^2 +\frac{D-3}{12(D+2)} c_2(R)
\right\}
(\dim \; R),
\label{ff3} \ea \nit
where we have used (\ref{j3}). Note also
\be \label{ff4} (M_t M_t)_{ab}=c_2(R)\delta_{ab}, \e \nit
compatibly with the second in result  (\ref{ff2}).

In the special case of $M=ad$, so that $L_M=L$, these results reproduce
those of Sec.~\ref{sect_l}, while (\ref{ff3}) gives rise to
\be \label{ff5}
{\rm Tr}\; L^4= \frac{D(D+27)}{12(D+2)} , \e \nit
since $c_2(ad)=1$. This can be confirmed correct using (\ref{f6}) and
results
from Sec.~\ref{sect_l}.

More specific applications of (\ref{ff3}) arise in
Sec.~\ref{sect_level3}, upon
identification of explicit expressions for the matrices of $R_2, R_3,
X_2$
etc.

%
\section{Simple tensor methods for $\g \in \cl{E}$}
%
\label{sect_simple}

\subsection{Second rank tensor decomposition}

Given a vector $v_i$ which transforms under $\g \in \cl{E}$ according to
its
adjoint representation, \ie\ an adjoint vector, we have, for $v_i v_j$
which transforms according to ${(ad \otimes ad)}_S$, the decomposition
into tensors
irreducible under $\g$:
\be \label{gg1} v_i v_j =\frac{1}{D} v_k v_k \delta_{ij}+d_{ija}x_a +
d_{ij\alpha} y_\alpha , \e \nit
where
\be \label{gg2} x_a =d_{ija} v_i v_j, \quad
y_\alpha  =d_{ij\alpha}v_i v_j. \e \nit
Here, to within normalisation, $d_{ija}, d_{ij\alpha}$ are
Clebsch-Gordan
coefficients, referred to often here as Clebsches for short, for
\bea ad \otimes ad & \rightarrow & R_2 \nonumber \\
ad \otimes ad & \rightarrow & R_3 . \label{gg3} \ea \nit
They are distinguished by virtue of having index sets of different
natures.
Our index conventions here are
\bea
i, j, k, \dots  \quad {\rm for} \quad & ad & \quad\in \{ 1, 2, \dots ,
\dim
\; \g\} \nonumber \\
a, b, c, \dots  \quad {\rm for} \quad & R_2 & \quad  \in \{ 1, 2, \dots
,
D_2=\dim R_2 \} \nonumber \\
\alpha, \beta, \gamma, \dots  \quad {\rm for} \quad  & R_3 & \quad  \in
\{ 1, 2, \dots , D_3=\dim R_3 \}, \label{gg4} \ea \nit
as well as
\be \label{gg5}
\mu, \nu, \rho, \dots \quad {\rm for} \quad X_2  \quad   \in \{ 1, 2,
\dots,
\Delta_2=\dim \; X_2 \}  , \e \nit
needed soon.

Eqs. (\ref{gg1}) and (\ref{gg2}) reflect the normalisations
\bea
d_{ija} d_{ijb} & = & \delta _{ab} \nonumber \\
d_{ij\alpha} d_{ij\beta} & = & \delta _{\alpha \beta} , \label{gg6} \ea
\nit
and the traceless properties
\be \label{gg7} d_{iia}=0, \quad d_{ii\alpha}=0. \e \nit
Also orthogonality of different sets of Clebsches gives
\be \label{gg8}
d_{ija} d_{ij\alpha} =0 . \e \nit

To bring $(ad \otimes ad)_A$ into the picture, let $\phi_i$ be a
fermionic
adjoint vector, \ie\ one with anticommuting components, \eg\
fermionic creation operators, as for $\alga_2$ in \cite{CdAMPB}. Then
the
analogue of (\ref{gg1}) is
\be \label{gg9}
\phi_i \phi_j = X_{ij} + c_{ijk}\psi_k , \e \nit
where
\be \label{gg10} \psi_i=c_{ijk}\phi_j \phi_k \e \nit
is the adjoint vector expected from
\be \label{gg11} (ad \otimes ad)_A \equiv ad + X_2 , \e \nit
and $X_{ij}$ is the tensor, clearly of dimension
\be \label{gg12}
\Delta_2=\fract{1}{2}D(D-3)=\dim \; X_2, \e \nit
associated with the second term of (\ref{gg11}). Also we may view the
$c_{ijk}$ as
Clebsches for $ad \otimes ad \rightarrow ad$.

\subsection{A projector view}

We obtain another useful view of (\ref{gg1}) by applying to $v_i v_j$
the
result, from (\ref{d7}),
\be \label{hh1}
I_S=P_1 +P_2 +P_3, \e \nit
where we write now
\bea (P_1)_{ij,kl} & = & \fract{1}{D} \delta_{ij}\delta_{kl} \nonumber
\\
(P_2)_{ij,kl} & = & d_{ija} d_{kla} \nonumber \\
(P_3)_{ij,kl} & = & d_{ij\alpha} d_{kl\alpha}. \label{hh2} \ea \nit
All the usual properties of projectors are satisfied:
${\rm Tr}\; P_1=1$ is trivial, $\; {\rm Tr}\; P_2=D_2$ and
${\rm Tr}\; P_3=D_3$ follow (\ref{gg6}), while
\be \label{hh3} P_2{}^2=P_2, \quad P_3{}^2=P_3, \quad P_2 P_3=0,
\quad P_1 P_3=0, \quad P_1 P_2=0 \e \nit
follow (\ref{gg6}), (\ref{gg7}) and (\ref{gg8}). Similarly we may apply
\be \label{hh4} I_A=P_{ad} + P_0 , \e \nit
to $\phi_i \phi_j$, where
\bea (P_{ad})_{ij,kl} & = & c_{ijt} c_{iklt} \nonumber \\
(P_0)_{ij,kl} & = & g_{ij\mu} g_{kl\mu}, \label{hh5} \ea \nit
where the Clebsches for $ad \otimes ad \rightarrow X_2$ are normalised,
like all the other Clebsches introduced so far, so that
\be \label{hh6} g_{ij\mu} g_{ij\nu}= \delta_{\mu \nu} . \e \nit
The application simply reproduces (\ref{gg9}) with
\be \label{hh7}
X_{ij}=(P_0)_{ij,kl} \phi_k \phi_l. \e \nit
We note also $ {\rm Tr}\; P_0=\delta_{\mu \mu}=\Delta_2=\dim \; X_2$,
and
the orthogonality relation
\be \label{hh8} c_{ijt} g_{ij\mu}=0 . \e \nit

Also alongside (\ref{gg5}) and (\ref{hh6}) we note the identities
\bea d_{ija} d_{kja} & = & \frac{D_2}{D} \delta_{ik} \nonumber \\
d_{ij\alpha} d_{kj\alpha} & = & \frac{D_3}{D} \delta_{ik} \nonumber \\
g_{ij\mu} g_{kj\mu} & = & \fract{1}{2} (D-3) \delta_{ik} . \ea \nit

\subsection{Two basic tensor identities}

From the equations
\bea I_S & = & P_1+P_2+P_3 \nonumber \\
LI_S & = & -P_1+\ell_2 P_2+\ell_3 P_3, \label{jj1} \ea \nit
we may eliminate $P_2$ and $P_3$ in turn, getting
\bea (L-\ell_2)I_S & = & -(1+\ell_2)P_1 +(\ell_3-\ell_2) P_3  \nonumber
\\
(L-\ell_3)I_S & = & -(1+\ell_3)P_1 +(\ell_2-\ell_3) P_2. \label{jj2} \ea
\nit
Taking matrix elements with the aid of (\ref{hh2}) gives
\be \label{jj3} (\ell_2-\ell_3) d_{ija} d_{pqa} =  -\fract{1}{2}(c_{ipt}
c_{jqt}+
c_{iqt} c_{jpt})  -\fract{1}{2} \ell_3(\delta_{ip} \delta_{jq}+
\delta_{iq} \delta_{jp})+\frac{1+\ell_3}{D}\delta_{ij} \delta_{pq},
\e \nit
plus a result for $d_{ij\alpha} d_{pq\alpha}$ obtained by interchange of
$\ell_2$ and $\ell_3$ in (\ref{jj3}).

It is a non-trivial but instructive task to verify that various
contractions of (\ref{jj3}) are identically satisfied; results from
Sec.~\ref{sect_quartic} are needed.

\subsection{Trilinear tensor identities}

Since we regard $c_{ijk}$ as defining the set of Clebsches for
\be \label{kk1} ad \otimes ad \rightarrow ad , \e \nit
eq. (\ref{x3}) can be regarded as an analogue of a result in the quantum
theory of angular momentum that defines a Racah coefficient.
There are many more identities of this sort. Appendix C provides a
little background from the quantum theory of angular momentum.

It is easy to contract (\ref{jj3}) with $d_{ija}$ and get
\be \label{kk2} c_{ikt} c_{tlj} d_{ija}= \ell_2 d_{kla} . \e \nit
There is a similar result for $d_{ij\alpha}$ obtained by replacing
$\ell_2$ on
the right of (\ref{kk2}) by $\ell_3$. There are many other pairs of
identities
related in this fashion; they should not need to be indicated explicitly
again.

To contract (\ref{jj3}) with a $c$-tensor and get
\be \label{kk3} d_{ija} d_{pqa} c_{ipt} = -\frac{\ell_2 D_2}{D} c_{pqs}
, \e \nit
is harder, and requires the use of (\ref{x3}) and results from
Sec.~\ref{sect_quartic}.

Also, using (\ref{hh4}), we get
\be \label{kk4} g_{ij\mu} g_{kl\mu} c_{iks} =- c_{jls} . \e \nit
Further, equipped with (\ref{kk2}), we can deduce
\be \label{kk5} d_{ija} d_{pqa} d_{ipb} = -\frac{11+D+\Delta}{2D\Delta}
d_{jqb} , \e \nit
where $\Delta$ is defined by (\ref{A1}).

A further consequence of (\ref{jj3}) is
\be \label{jj4} d_{(ij}{}^a d_{pq)a}= \frac{2D_2}{D(D+2)}
\delta_{(ij} \delta_{pq)}, \e \nit
in which the round brackets denote symmetrisation over the enclosed at
unit
weight. The first index $a$ is raised, without any metric significance,
just
to take it outside the round brackets. This result, (\ref{jj4}), can be
used
to give an independent derivation of (\ref{kk5}), with the aid of the
result
in (\ref{A3}) for $D_2$,

Obviously there are more results of the type here treated, of increasing
complication. We have shown how one might work towards them if and when
the
need to do so arises.

%
\section{Matrices and associated tensors}
%
\label{sect_matrices}

\subsection{The basis set}

To gain full control of the formalism, systematically identifying all
the
tensors of importance as they arise and determining their essential
properties,
it is useful to introduce a complete set of basis matrices for
$\hom_\R(\cl{V}_{ad},\cl{V}_{ad})$, where $\cl{V}_{ad}$ is the vector
space in
which the irrep $ad$ of $\g \in \cl{E}$ acts.

The basis
\be \label{X1}
M_A, \quad A \in \{ 1, 2, \dots, D^2 \}, \quad D= \dim \; \g , \e \nit
consists of matrices
\be \label{X2}
 (\frac{1}{D} I, \; D_a, \; Y_\alpha), (F_i, \; G_\mu). \e \nit
The first parenthesis contains a total of $1+D_2+D_3=\frac{1}{2}D(D+1)$
symmetric matrices defined by
\be \label{X3}
I_{ij}=\delta_{ij}, \quad (D_a)_{ij}= -d_{ija}, \quad (Y_\alpha)_{ij}
=-d_{ij\alpha}. \e \nit
The second parenthesis in (\ref{X2}) contains
$D+\Delta_2=\frac{1}{2}D(D-1)$
antisymmetric matrices defined
by
\be \label{X4}
(F_i)_{jk}=-ic_{ijk}, \quad (G_\mu)_{ij}=-ig_{ij\mu} . \e \nit
These definitions are all given in terms of tensors already introduced
in
Sec.~\ref{sect_simple}.

The matrices $M_A$ all hermitian, and possess the trace properties
\be \label{X5}
{\rm Tr}\; M_A=0, \quad {\rm Tr}\; (M_A M_B)= \delta_{AB} . \e \nit
By expanding symmetric $A \in \hom_\R(\cl{V}_{ad},\cl{V}_{ad})$ with
respect to
our basis
\be \label{X6} A=aI+u_a D_a +v_\alpha Y_\alpha \e \nit
we obtain a completeness relation
\be \label{X7}
d_{ija} d_{kla} +d_{ij\alpha} d_{kl\alpha}
+\fract{1}{D}\delta_{ij}\delta_{kl}
=\fract{1}{2}(\delta_{ik}\delta_{jl}+\delta_{il}\delta_{jk}) , \e \nit
the compatibility of which with (\ref{jj4}) and its analogue involving
$D_3$
can be checked.

Similarly we have
\be \label{X8}
c_{ikt} c_{jlt} +g_{ik\mu} g_{jl\mu}=
\fract{1}{2}(\delta_{ik}\delta_{jl}-\delta_{il}\delta_{jk}), \e \nit
which contains the same information as (\ref{hh4}).

\subsection{The product $F_i F_j$}

By considering the action of $I=I_A+I_S$ on $F_i F_j$ we find
\be \label{ll1} F_i F_j=\frac{1}{D} \delta_{ij}+ \fract{1}{2}i c_{ijk}
F_k
+\ell_2 d_{ija} D_a +\ell_3 d_{ij\alpha} Y_\alpha , \e \nit
with no term in $G_\mu$ because of the closure of the Lie algebra $\g$.

Here we have used facts like
\be \label{ll2} {\rm Tr}\; (F_i F_j D_a)=c_{piq} c_{qjr} d_{rpa}= \ell_2
d_{ija},
\e \nit evaluated using (\ref{kk2}). The factors $i$ needed for the
hermiticity
of the $F$-matrices accounts for the minus sign in the definition
(\ref{X3}).

\subsection{The full set of product laws}

It is necessary to be prepared to contemplate all the product laws
within
$M_A M_B$ and all the tensors that arise in them. The full list is
\bea
 F_i F_j & = & \frac{1}{D} \delta_{ij}+ \fract{1}{2} i c_{ijk} F_k
+\ell_2 d_{ija} D_a +\ell_3 d_{ij\alpha} Y_\alpha \label{P1} \\
F_i D_a & = & \fract{1}{2}i m_{iab} D_b+\ell_2 d_{ija} F_j +
d_{ia\mu}G_\mu
\label{P2} \\
F_i Y_\alpha & = & \fract{1}{2}i m_{i\alpha \beta} Y_\beta+\ell_3
d_{ij\alpha} F_j + d_{i\alpha \mu}G_\mu
\label{P3} \\
F_i G_\mu & = & \fract{1}{2}i g_{i\mu \nu} G_\nu+
d_{i\alpha\mu} Y_\alpha+d_{ia\mu} D_a
\label{P4} \\
D_a D_b & = & \frac{1}{D} \delta_{ab}I+ \fract{1}{2}i m_{iab} F_i+
\fract{1}{2}i m_{\mu ab} G_\mu
+d_{abc} D_c+ d_{ab\alpha} Y_\alpha \label{P5} \\
D_a Y_\alpha & = & \fract{1}{2}i m_{a\alpha \mu} G_\mu+ d_{ab \alpha}
D_b+
d_{a\alpha \beta} Y_\beta \label{P6} \\
G_\mu D_a & = &  \fract{1}{2}i m_{\mu ab} D_b + \fract{1}{2}i m_{a\alpha
\mu}
Y_\alpha +  d_{ia\mu} F_i+ d_{a \mu \nu} G_\nu \label{P7} \\
Y_\alpha Y_\beta & = & \frac{1}{D} \delta_{\alpha \beta} +
 \fract{1}{2}i m_{i\alpha \beta} F_i + \fract{1}{2}i m_{\mu \alpha
\beta}
G_\mu + d_{\alpha \beta \gamma} Y_\gamma +d_{a\alpha \beta} D_a
\label{P8} \\
Y_\alpha G_\mu & = & \fract{1}{2}i m_{a\alpha \mu} D_a+
\fract{1}{2}i m_{\mu \alpha \beta} Y_\beta+d_{i\alpha \mu} F_i+
d_{\alpha \mu \nu} G_\nu \label{P9} \\
G_\mu G_\nu & = &  \frac{1}{D} \delta_{\mu \nu}+\fract{1}{2}i g_{i\mu
\nu} F_i
+\fract{1}{2}i g_{\mu \nu \rho} G_\rho+d_{a \mu \nu} D_a+
d_{\alpha \mu \nu} Y_\alpha. \label{P10} \ea

Thus we need to consider $10$ products with $39$ terms,
involving $4$ Kronecker deltas and $18$ isotropic third rank tensors.
Again, tensors named by the same letter, but
carrying distinct types of index sets, are to be regarded
as distinct tensors.
To understand fully the detail contained in (\ref{P1}--\ref{P10}), it is
best
first to consider commutators and anticommutators separately. Thus, \eg,
${[} F_i \, , \, D_a {]}$ is symmetric and lies in ${\rm span}_\R(
I, D_a, Y_\alpha)$, while $\{ F_i \, , \, D_a \}$ is antisymmetric and
lies in
 ${\rm span}_\R(F_i, G_\mu)$. The question then is: why are there only
three terms in (\ref{P2}). To answer we note some direct product results

\bea
 ad \otimes X_2 & =& ad + X_2+R_2+R_3+R_5+R_6+X_3+R_9 \label{h3} \\
 ad \otimes R_2 & =& ad + X_2+R_2+R_4+R_5+R_6 \label{h4} \\
 ad \otimes R_3 & =& ad + X_2+R_3+R_6+R_8+R_9. \label{h5} \ea \nit
These results, all of which are relevant at this point even though we
may not
make this completely explicit, bring in
six families of irreps which have not so far been mentioned.
These are defined by their dimensions
for $\g_2, \f_4, \alge_{6-8}$ in that order, as follows
\bea R_4 & = & 7, 273, 650, 1463, - \label{h6} \\
 R_5 & = & 64, 4096, 11648, 40755, 147250 \label{h7} \\
 R_6 & = & 189, 10829, 34749, 152152, 779247 \label{h8} \\
 X_3 (=R_7) & = & 182, 19448, 70070, 365750, 2450240 \label{h9} \\
 R_8 & = & 273, 12376, 43758, 238602, 1763125 \label{h10} \\
 R_9 & = & 448, 29172, 105600, 573440, 4096000. \label{h11} \ea \nit

The point here is that the result (\ref{h4}) relevant to (\ref{P2})
contains
only three terms relevant to our basis matrices $M_A$ namely the first
three,
which correspond to the $F, G, D$ terms of (\ref{P2}). There is no $Y$
term
in (\ref{P2}) because $R_3$ does not occur in (\ref{h4}).
Tracing with $F_k$ accounts for the coefficient $\ell_2 d_{ija}$, while
the other
two allowed terms necessitate the introduction of two new tensors.
Eqs. (\ref{h3}--\ref{h5}) relate similarly to (\ref{P4}--\ref{P2}).
And so on
one proceeds. The repetition in later products of tensors introduced in
earlier
products is accounted for by requiring consistency under tracing.
Various tensors have obvious symmetry or antisymmetry properties. Also
the
order of various matrices of different types within trace definitions
should
be irrelevant, to within a sign; \eg\ it can be proved that the three
traces
\be \label{P21}  {\rm Tr}\; (F_i D_a G_\mu)=
{\rm Tr}\; (F_i G_\mu D_a)={\rm Tr}\; (G_\mu F_i D_a)=d_{ia\mu}, \e \nit
are mutually consistent.

We draw attention to the absence of an $F$-term in (\ref{P4}). This
follows
from Lie algebra closure. Another view of this states that
\be \label{P101} {\rm Tr} \; (F_i F_j G_\mu)=0. \e \nit
This in turn gives an identity that can be proved, using the Jacobi
identity for the structure constants, in exactly the same way as
(\ref{x3})
was proved
\be \label{P102} c_{piq} c_{qjr} g_{rp\mu}=0. \e

\subsection{Tensors as Clebsches}
\label{sect_clebsh}

There are in the product laws (\ref{P1}--\ref{P10}) four distinct
Kronecker
deltas associated trivially with
\be \label{mm1}
R_1=1 \in ad \otimes ad, \quad R_2 \otimes R_2, \quad R_3 \otimes R_3,
\quad
X_2 \otimes X_2 . \e \nit

\begin{table}
\begin{center}
\begin{tabular}{|c|c|c|} \hline
   Notation & Symmetries & Triad of irreps   \\ \hline \hline
$c_{ijk}$ & TA & $ad \otimes ad \rightarrow ad$ \\
$d_{ija}$ & S & $ad \otimes ad \rightarrow R_2$ \\
$d_{ij\alpha}$ & S  & $ad \otimes ad \rightarrow R_3$ \\
$m_{iab}$ & A & $ad \otimes R_2 \rightarrow R_2$ \\
$m_{i\alpha \beta}$ & A & $ad \otimes R_3  \rightarrow R_3$ \\
$d_{ia\mu}$ & & $ad \otimes R_2 \rightarrow X_2$ \\
$d_{i\alpha \mu}$ & & $ad \otimes R_3 \rightarrow X_2$ \\
$m_{\mu ab}$ & A & $R_2 \otimes R_2 \rightarrow X_2$ \\
$d_{abc}$ & TS & $R_2 \otimes R_2 \rightarrow R_2$ \\
$d_{ab\alpha}$ & S & $R_2 \otimes R_2 \rightarrow R_3$ \\
$m_{\mu \alpha \beta}$ & A & $R_3 \otimes R_3 \rightarrow X_2$ \\
$d_{\alpha \beta \gamma}$ & TS & $R_3 \otimes R_3 \rightarrow R_3$ \\
$d_{\alpha \beta a}$ & S & $R_3 \otimes  R_3 \rightarrow  R_2$ \\
$g_{i\mu \nu}$ & A & $X_2 \otimes X_2 \rightarrow  ad$ \\
$m_{a\alpha \mu}$ & & $R_2 \otimes R_3 \rightarrow X_2$  \\
$d_{a \mu \nu}$ & S & $X_2 \otimes X_2 \rightarrow R_2$ \\
$d_{\alpha \mu \nu}$ & S & $X_2 \otimes X_2  \rightarrow  R_3$ \\
$g_{\mu \nu \rho}$ & TA  & $X_2 \otimes X_2 \rightarrow X_2$ \\  \hline
\end{tabular}
\end{center}
\caption{\label{tab_tensors}Third rank isotropic tensors}
\end{table}

For the third rank isotropic tensors we have drawn up
Table~\ref{tab_tensors}. It indicates symmetry properties with respect
to interchange of indices of the same type, with the letter $T$
standing for totally. The table also specifies the triad of irreps for
which each tensors provides a set of Clebsches. We emphasise that only
the terms $ad, R_2, R_3, X_2$ on the right sides of
(\ref{h3}--\ref{h5}) are relevant at present, but see
Sec.~\ref{sect_level3}.

\subsection{Jacobi identities and matrix irreps of $\g \in \cl{E}$}
\label{sect_jacobi}

The results, from (\ref{P2}--\ref{P4}),
\bea {[} F_i \; , \; D_a {]} & = & im_{iab} D_b \nonumber \\
{[} F_i \; , \; Y_\alpha {]} & = & im_{i\alpha \beta} Y_\beta \nonumber
\\
{[} F_i \; , \; G_\mu{]} & = & ig_{i\mu \nu} G_\nu, \label{U1} \ea \nit
imply that $D_a, \; Y_\alpha, \; G_\mu$ transform under $\g \in \cl{E}$
according to the irreps $R_2, \; R_3, \; X_2$ of $\g \in \cl{E}$. The
tensors
that appear on the right side of (\ref{U1}) are very important ones.
To see this, we use Jacobi identities of the sort $F, \; F, \; X$ for
$X=
D, \; Y, \; G$ in turn. The first one translates into
\bea
 {[} M_i \; , \; M_j {]} & = & ic_{ijk} M_k , \label{U2} \ea \nit
where the matrices $M_i$ are defined by
\be \label{U3} (M_i)_{ab}=-im_{iab} . \e \nit
Thus $X_i \mapsto M_i$ defines the $D_2 \times D_2$ matrices of the
irrep
$R_2$ of $\g$.

Similarly
\bea (N_1)_{\alpha \beta} & = & -im_{i\alpha \beta} \nonumber \\
(G_i)_{\mu \nu} & = & -ig_{i\mu \nu} , \label{U4} \ea \nit
defines the matrix irreps $X_i \mapsto N_i$ and to $G_i$ for the irreps
$R_2$ and $X_2$ of $\g \in \cl{E}$.

\subsection{Eigenvalues of ${\cl{C}}^{(2)}$ for $X_i \mapsto M_i, \;
N_i, \;
G_i$}

It is easiest in the case of $G_i$ to show that the definition of
Sec.~\ref{sect_jacobi} is consistent with the knowledge, already to
hand, that
\be \label{V1} G_i G_i =2I . \e \nit
Thus we note the results
\bea (G_i G_i)_{\mu \rho} & = & g_{i\mu \nu} g_{i \nu \rho} \nonumber \\
\fract{1}{2} i g_{m\mu \nu} & = & {\rm Tr}\; (G_\mu G_\nu F_m)
=-g_{jk \mu} g_{kl\nu} c_{jlm}.  \label{V2} \ea \nit
We insert the second one, not directly into the first, but rather into
\be \label{V3} g_{m\mu \nu} g_{n\mu \nu}. \e \nit
Then use of (\ref{x3}, \ref{P102}, \ref{a4}) and (\ref{hh6})  enables
the
proof that
\be \label{V4} g_{m\mu \nu} g_{n\mu \nu}=(D-3) \delta _{mn} . \e \nit
This is tantamount to proving
\be \label{V5} g_{m\mu \nu} g_{m\nu \rho}= 2\delta_{\mu \rho} , \e \nit
which is as required.

The same method can be applied to showing that
\bea
(M_i M_i)_{ac} & = & m_{iab} m_{ibc}= 2(1+\ell_2)\delta_{ac} \nonumber
\\
(N_i N_i)_{\alpha \gamma} & = & m_{i\alpha \beta} m_{i\beta \gamma}=
2(1+\ell_3)\delta_{\alpha \gamma}. \label{V6} \ea \nit
The first of these requires (\ref{jj3}), (\ref{kk3}) etc., and emerges
upon use of formulas form Sec.~\ref{sect_quartic}.

In (\ref{V5}) and (\ref{V6}), the required eigenvalues of $\cl{C}^{(2)}$
are seen explicitly on the right sides.

%
\section{Towards $ad \otimes ad \otimes ad$}
%
\label{sect_level3}

\subsection{$\dim \; X_3$ and $c_2(X_3)$}
\label{sect_x3}

The result
\be \label{au1} (ad \otimes ad \otimes ad)_A \equiv R_1+R_2+R_3+X_2+X_3,
\e \nit is known from \cite{CdM} on the basis of \cite{LiE}. It can be
proved,
as in \cite{mp2002}, using methods based on the Molien function
\cite{mol,sw}, a method capable \cite{mp2002}
of treating higher
$ad^{\wedge r}$.

Its importance resides in the fact that all but the last
family of irreps of $\g \in \cl{E}$ have been treated fully already at
the
level
of $ad \otimes ad$. Accordingly (\ref{au1}) gives us an easy passage to
the
treatment of the family $X_3$.

We define the operator $M$ via
\bea \cl{C}^{(2)} & = & (X_1+X_2+X_3)^2= 3+2M \nonumber \\
M & = & X_1.X_2+X_2.X_3+X_3.X_1, \label{au2} \ea \nit
in which all the $X_i=ad_i$. We define projectors onto the eigenspaces
of
$\cl{C}^{(2)}$, and hence of $M$, so that
\bea I_A & = & P_1+P_2+P_3+P_0+P_7 \nonumber \\
\fract{1}{6} D(D-1)(D-2) & = & (1+D_2+D_3)+ \Delta_2+ \Delta_3,
\label{au3}
\ea \nit
where $\Delta_{2,3}=\dim \; X_{2,3}$. Hence
\be \label{au4} \Delta_3=\fract{1}{6} D(D-1)(D-8) . \e \nit

Next we apply $M=\fract{1}{2}(\cl{C}^{(2)}-3)$ to the first entry of
(\ref{au3}).This gives \be \label{au5}
{\rm Tr}\; (M I_A)= -\fract{3}{2}+(\ell_2-\fract{1}{2})D_2
+(\ell_3-\fract{1}{2})D_3 -\fract{1}{2}\Delta_2+m\Delta_3 , \e \nit
where $m$ is the eigenvalue of $M$ for $X_3$.
One calculates the left side directly getting $D-\fract{1}{2}D^2$.
All the other quantities in (\ref{au5}) are known as functions of
$D=\dim \g$.
Hence, with the aid of (\ref{g1}),
we find that $m=0$, so that $\cl{C}^{(2)}$ has eigenvalue
\be \label{au6} c_2(X_3)=3, \e \nit
completing algebraic derivation of the expected result.

\subsection{Trace equations for $ad \otimes ad \otimes ad$}

The approach here is based on the results (\ref{h3} -- \ref{h5}), and
depends
on the fact that formulas for $c_2(R)$ and $\dim \; R$ are known for
\be \label{bv1} R \in \quad \{ad, R_2, R_3, X_2, X_3(\equiv R_7) \} . \e
\nit
It will be seen soon that the fact that $X_3$ has been treated (in
Sec.~\ref{sect_x3}) is crucial, enabling us to deduce corresponding
results for
\be \label{bv2} R \in \quad \{ R_4, R_5, R_6, R_8, R_9 \} . \e \nit
we begin by calculating
\be \label{bv3} {\rm Tr} \; \cl{C}^r, \quad r=0, 1, 2, 3 \; , \e \nit
where
\be \label{bv4} \cl{C}= \cl{C}^{(2)}{}_{ad \otimes R}, \quad R
=X_2, R_2, R_3. \e \nit
We have
\be \label{bv5} \cl{C}^{(2)}{}_{ad \otimes R}=(ad_t+M_t)(ad_t+M_t)
=1+c_2(R)+L_M, \e \nit
where $X_i \mapsto M_i$ defines the matrices of $R$, and $L_M$ is as
defined in
(\ref{ff1}) by $L_M=ad_t M_t$, so that ${\rm Tr}\;(L_M)^r$ is known,
from the work
of Sec.~\ref{sect_c4}, for $r=0, 1, 2, 3$. It is known for $m=4$ also
but this is not needed now. Thus we obtain
\bea {\rm Tr}\; I & = & DD(R) \nonumber \\
{\rm Tr}\; \cl{C} & = & DD(R)(1+c_2(R))  \nonumber \\
{\rm Tr}\; \cl{C}^2 & = & DD(R)(1+c_2(R))^2 +4c_2(R) D(R)  \nonumber \\
{\rm Tr}\; \cl{C}^3 & = & DD(R)(1+c_2(R))^3
+12c_2(R) D(R)(1+c_2(R))
-2c_2(R) D(R). \label{bv6} \ea \nit

We now outline how, by reference to (\ref{h3}) and (\ref{h4}), we
can evaluate $c_2(R)$ and $D(R)$ for $R_9$ and $R_4$. A similar method
applied
to (\ref{h3}) and (\ref{h5}) can be used to treat $R_5$ and $R_8$,
leaving the
easy final step of handling $R_6$ to complete the job.

From (\ref{h3}) and (\ref{h4}) we get
\bea I & = & P_{ad} +P_0+P_2+P_3+P_5+P_6+P_7+P_9 \nonumber \\
I & = & P_{ad} +P_0+P_2+P_4+P_5+P_6 . \label{bv7} \ea \nit
We have not distinguished the different unit operators. Taking
the traces of these equations and subtracting allows much cancellation
and
gives
\be \label{bv8} D_9-D_4= D(\Delta_2-D_2) -(D_3+D_7)= f_0(D) , \e \nit
where $D_9=\dim \; R_9,  D_4=\dim \; R_4$. Next acting on (\ref{bv7})
with
the appropriate Casimirs $\cl{C}$, taking traces using (\ref{bv6}),
and subtracting, gives
\be \label{bv9}  c_9 D_9-c_4D_4= f_1(D) , \e \nit
where
\be \label{bv10} f_1(D)=3D\Delta_2-DD_2(1+c_2)-(c_3D_3+3D_7), \e \nit
and $c_r=c_2(R_r), r \in \{ 9, 4, 2, 3, 7 \}$ with $c_7=3$. Similarly
using
the square and the cube of $\cl{C}$ we complete the derivation of the
set
of four equations
\be \label{bv11} c_9{}^r D_9-c_4{}^r=f_r(D) , \quad r=0, 1, 2, 3,
\e \nit
where we have not displayed expressions for $f_2(D)$ or $f_3(D)$.
The method of Sec.~\ref{sect_ells} (matrixwise elimination of $D_9$ and
$D_4$)
now immediately yields
\bea c_9+c_4 & = & \frac{f_0 f_3-f_1 f_2}{f_0 f_2-f_1{}^2} \nonumber \\
c_9 c_4 & = & \frac{f_1 f_3-f_2{}^2}{f_0 f_2-f_1{}^2}. \label{bv12} \ea
\nit
It is obvious how to assign the two solutions of these equations
appropriately
to the correct families, $R_4, \; R_9$.
The explicit evaluation of the right sides of (\ref{bv12}) is a task
best left to MAPLE. Because of non-rational dependence on $D$, it
is better to work in terms of $m$, related to $D$ by (\ref{K5}) of
Appendix A.
However the results are
already known: \cite{CdM}, where \cite{LiE} was employed.  Since,
in confirming them, we have used a
different parametrisation from that of \cite{CdM}, we quote
\bea c_9=c_2(R_9)=\frac{3m+7}{m+2} & , & c_4=c_2(R_4)=\frac{2(m+1)}{m+2}
\nonumber \\
c_5=c_2(R_5)=\frac{5m+8}{2(m+2)} & , & c_8=c_2(R_8)=\frac{3m+8}{m+2}
\nonumber \\
c_6=c_2(R_6) = \frac{8}{3} & . & \label{bv13} \ea
Further, for convenience of readers, we have listed the expressions in
terms of
$m$ for the dimensions of
\be \label{bv124} X_2=R_0, R_2, \dots, R_6, X_3=R_7, R_8, R_9  \e \nit
in Appendix B. Once $c_9$ and $c_4$ have been found it is a simple
matter to
use (\ref{bv8}--\ref{bv9}) to reach $D_9$ and $D_4$, etc.

%
\section{The case of $\algd_4$, and of its quartic Casimirs}
%
\label{sect_d4}

\subsection{$ad \otimes ad$}

The versions of (\ref{I3}) and (\ref{I4}) that apply to $\algd_4$  read
as
\bea
(ad \otimes ad)_A & = & ad +350 \label{T0} \\
(ad \otimes ad)_S & = & 1+\{ 35+35+35 \}+300 \nonumber \\
& = & (0,0,0,0)+\{ (2,0,0,0)+(0,0,2,0)+(0,0,0,2) \} +(0,2,0,0).
\label{T1}
\ea \nit
The irrep $350$ here agrees with (\ref{I5}) for $\dim \; \algd_4=D=28$,
but, in the role of $R_2$ in (\ref{I4}), (\ref{T1}) suggests the direct
sum
of three
inequivalent irreps of $\algd_4$. These three irreps, whose Dynkin
labels are
given explicitly above, are a set  of three related by triality, all of
which
share the eigenvalue $\fract{4}{3}$ of $\cl{C}^{(2)}$ for $\algd_4$. It
is the
latter fact that enables their direct sum, viewed as a single entity,
to fulfill exactly the role of $R_2$ in the general discussion that
applies to
other members of $\g \in \cl{F}$.

Since the parameter $\Delta$ of (\ref{A1}) has the value $3$ for
$\algd_4$,
we get $D_2=105$, correctly, and the expected values of $D_3,
\ell_2,\ell_3$ for
$\algd_4$ by inserting
$\Delta = 3$ into the results of Sec.~\ref{sect_quartic}.

The discussion of the situation surrounding irreps of $\algd_4$ related
by
triality, such as the $35$'s in (\ref{T1}), can be refined by
consideration
of irreps of the group obtained by extending the group $SO(8)$ by the
group of
automorphisms of its Dynkin diagram. Here we merely refer to \cite{CdM}
for this and similar considerations for $\alga_2$ and $\alge_6$.

\subsection{The quartic Casimirs of $\algd_4$}

Sec.~\ref{sect_c4} explains why the exceptional Lie algebras $\cl{E}
\subset \cl{F}$
do not possess a primitive quartic Casimir operator. Since
$\algd_4=so(8)
\subset \cl{F}$
has two independent primitive quartic Casimir operators, it might seem
that
$\algd_4$ fails to conform fully to its implied status within $\cl{F}$.
We show
next that is not the case, showing explicitly exactly how it conforms.

The projector $P_2$ that projects onto the $R_2$ subspace of
$(ad \otimes ad)_S$ is given by (\ref{hh2}) for all $\g \in \cl{F}$ in
the form
\be \label{T100} (P_2)_{ij,pq}= d_{ija} d_{pqa}, \e \nit
and a view of the $d_{ija}$ as a set of Clebsches for $ad \otimes ad
\rightarrow R_2$ was indicated in Sec.~\ref{sect_clebsh}.
Since
\be \label{T101}
P_2=\frac{(L+1)(L-\ell_3)I_S}{(\ell_2+1)(\ell_2-\ell_3)}, \e \nit
we find, with a temporary abbreviation $g(D)$ for the denominator of the
right side of (\ref{T101}),
\be \label{T102}
g(D) d_{ija}d_{pqa}=\fract{1}{2} c_{ikt} c_{ilt}
(c_{kps} c_{lqs}+c_{kqs} c_{lps})-\fract{1}{2} (1-\ell_3)(c_{ipt}
c_{jqt}+
c_{iqt} c_{jpt})+\fract{1}{2} \ell_3 (\delta _{ip} \delta_{jq}+
\delta_{iq} \delta_{jp}). \e \nit
If we now define a quartic Casimir invariant for the vector $v_i$
according to
\be \label{T103} Q^{(4)}= d_{ija}d_{pqa} v_i v_j v_p v_q ,
\e \nit
we get
\be \label{T104} g(D)Q^{(4)}=c_{ikt} c_{tjl} c_{lqs} c_{spk} v_i v_j v_p
v_q
-\ell_3v_k v_k v_l v_l, \e \nit
which shows the definition (\ref{T103}) is a satisfactory
alternative to that of $\cl{C}^{(4)}$ used in (\ref{j1}). It further
reduces,
as $\cl{C}^{(4)}$ itself reduced using (\ref{f7}), to a multiple of the
square of the
quadratic invariant $v_k v_k$. We do not exhibit the result as the
multiple
does not simplify into a nice enough form.

Putting $D=28$ and $\ell_3=\fract{1}{6}$ naively into our result for
$Q^{(4)}$, we
find
\be \label{T105} Q^{(4)} =\fract{1}{2}v_k v_k v_l v_l . \e \nit

Using the notation $x_a=d_{ija} v_i v_j$ of (\ref{gg2}), we write this
as
\be \label{T106}
2Q^{(4)}=x_a x_a. \e \nit
As noted for the exceptionals, this is the whole story;
there is one irreducible vector $x_a$, and one equation, \eg\
(\ref{T105})
which means that the square $x_a x_a$
does not define a primitive quartic Casimir.
For $\algd_4$ the difference from other $\g \in \cl{F}$ lies in the
reducibility
of the
representation $R_2$ for $\algd_4$. In fact, the projector $P_2$ is the
sum of
three orthogonal projectors. Put otherwise, there are three {\it
orthogonal}
sets of Clebsch-Gordan coefficients for $35$'s belonging to $(ad \otimes
ad)_S$
and three pairwise  orthogonal $35$ component entities
\be \label{R4} w^{(r)}{}_a= k_r d^{(r)}{}_{ija} v_i v_j , \e \nit
in which the $k_r$ are constants,
such that
\be \label{R5} x_a=w^{(1)}{}_a+w^{(2)}{}_a+w^{(3)}{}_a. \e \nit
Now $x_a x_a$ itself is not itself primitive. But, since the $w^{(r)}$
are
orthogonal,
\be \label{R6} x_a x_a= w^{(1)}{}_a w^{(1)}{}_a+
w^{(2)}{}_a w^{(2)}{}_a+w^{(3)}{}_a w^{(3)}{}_a, \e \nit
and this leaves two linear
combinations of the three squares, which can serve as independent and
primitive quartic invariants. This places the known situation for
$\algd_4$
correctly within, and not superficially outside, the family context.
\vskip 10pt

\begin{flushleft} {{\bf Acknowledgements}}
\end{flushleft}
The research of AJM is supported in
part by PPARC. HP is grateful to Emmanuel College, Cambridge, for a
Research
Fellowship. We thank Bruce Westbury for stimulating discussions, and for
generously providing us with copies of manuscripts of his research work,
including a preliminary version of \cite{bw}.

\vskip 15pt
\begin{flushleft} {{\bf Appendix A: $\quad$ Other parametrisations}}
\end{flushleft} \vskip 3pt
In our work we have chosen to use the $D=\dim \; \g$ as the parameter
in formulas such as (\ref{I4}) for dimensions $\dim \; R$, or
eigenvalues
$c_2(R)$ of $\cl{C}^{(2)}$ with
\be \label{K1}
 D=3,8, 14, 28, 52, 78, 133, 248 , \e \nit
for $\g \in \cl{F}$. Other workers in the general area have made
different choices. In \cite{del} and \cite{CdM} one finds
\be \label{K2} \alpha= \fract{1}{2}, \fract{1}{3},  \fract{1}{4},
\fract{1}{6},
\fract{1}{9}, \fract{1}{12}, \fract{1}{18}, \fract{1}{30}. \e \nit
This has the significance that $\alpha$ is the inverse of the dual
Coxeter
number $h^{\vee}$ for each $\g$, \cite{fss} p37.
The choice has a natural interpretation also in terms of our work:
\be \label{K3} \alpha =\ell_3= \fract{1}{2}c_2(R_3)-1. \e \nit
Here $\ell_3$ denotes the eigenvalue for $R_3$ of the $L$-operator
used in Sec.~\ref{sect_level2} in the analysis of $ad \otimes ad $.

In recent studies\cite{lm,bw}, one meets the parameter $m$
with values
\be \label{K4} m=-\fract{4}{3},-1, -\fract{2}{3}, 0, 1, 2 ,4, 8 \e \nit
related to $D$ via
\be \label{K5} D=\frac{2(3m+7)(5m+8)}{m+4}, \e \nit
and to $\alpha =\ell_3$ via
\be \label{K6} \alpha =\ell_3= \frac{1}{3(m+2)}, \e \nit
so that $h^{\vee}=3(m+2)$.

For the Lie algebras of the last line of the Freudenthal magic square
itself
there is the further observation that $m$ is equal to the dimension of
the
division algebra involved in the Freudenthal construction of each one.

One other thought: suppose one solves (\ref{K5}) for $m$ in terms of
$D$. Of
the two roots of the quadratic equation in question here, one is $m$ and
has
values related to $\ell_3$ by (\ref{K6}). The other root has different
values, $m^\prime$, say, such that
\be \label{K7} \ell_2= \frac{1}{3(m^\prime +2)}, \e \nit
where $\ell_2$ denotes the eigenvalue for $R_2$ of the $L$-operator used
in
Sec.~\ref{sect_simple}. Comparison of (\ref{K6}) and (\ref{K7})
reveals a close relationship to the involution $*$ used in \cite{CdM}.

We also note the parameter $\Delta=\Delta(D)$ of (\ref{A1}), and its
role, see Sec.~\ref{sect_explicit}, in formulas not dependent linearly
upon
$D$. Also $\Delta \mapsto -\Delta$ corresponds to the star involution
of \cite{CdM}.

\vskip 10pt
\begin{flushleft} {{\bf Appendix B: Dimension formulas in terms of $m$}}
\end{flushleft} \vskip 5pt

As mentioned above, many formulas are in essentially their simplest form
when
written in terms of $m$ rather than $D=\dim \g$, especially ones which
involve
the quantity $\Delta$ of (\ref{A1}), a rational function of $m$ but not
of $D$.
This applies to many dimension formulas. We have
\vfill \eject

\be \label{df0} D= \dim \g = {{2(5m+8)(3m+7)} \over {(m+4)}}, \e \nit
\be \label{df1}
\dim \; R_0 \equiv \dim \; X_2= {{5(5m+8)(3m+7)(3m+4)(2m+5)} \over
{(m+4)^2}},
\e \nit
\be \label{df2}
\dim \; R_2={{90(3m+7)(3m+4)(m+2)^2} \over {(m+6)(m+4)^2}}, \e \nit
\be \label{df3}
\dim \; R_3={{45(5m+8)(2m+5)(m+2)^2} \over {(m+6)(m+4)}}, \e \nit
\be \label{df4}
\dim \; R_4={{5(5m+8)(3m+7)(3m+4)(2m+3)(m+2)^(8-m)} \over
{(m+6)(m+5)(m+4)^3}}, \e \nit
\be \label{df5}
\dim \; R_5={{5120(3m+7)(2m+5)(2m+3)(m+2)^2(m+1)} \over
{(m+8)(m+6)(m+4)^3}},
\e \nit
\be \label{df6}
\dim \; R_6=
{{27(5m+12)(5m+8)(3m+7)(3m+4)(2m+5)(2m+3)} \over {(m+9)(m+5)(m+4)^2}},
\e \nit
\be \label{df7}
\dim \; R_7={{10(5m+12)(5m+8)(3m+8)(3m+7)(2m+3)(m+1)} \over {(m+4)^3}},
\e \nit
\be \label{df8}
\dim \; R_8={{10(5m+12)(5m+8)(3m+11)(3m+7)(2m+5)(m+2)^2} \over
{(m+8)(m+6)(m+4)^2}}, \e \nit
\be \label{df9}
\dim \; R_9={{40(5m+12)(5m+8)(3m+8)(3m+4)(m+2)^2} \over
{(m+6)(m+5)(m+4)}}.
\e \nit
\pagebreak
\begin{flushleft} {{\bf Appendix C: $\quad$ Racah coefficients}}
\end{flushleft} \vskip 5pt

For more details the reader may refer to textbooks devoted to the
quantum
theory of angular momentum, or to the valuable reprint volume
\cite{BvD}.

Racah coefficients arise in the comparison of different ways of coupling
three angular momenta to define the total angular momentum. One way of
presenting the definition in terms of angular momentum Clebsches is
\bea
\sum_m \langle j_1 j_6 m_1 m_6 | j_5 m_5 \rangle
\langle j_2 j_4 m_2 m_4 | j_6 m_6 \rangle
\langle j_3 j_4 m_3 m_4 | j_5 m_5 \rangle = & {} & \nonumber \\
\sqrt{(2j_3+1)(2j_6+1)}  W(j_1 j_2 j_5 j_4, j_3 j_6) & {} &
\langle j_1 j_2 m_1 m_2 | j_3 m_3 \rangle , \label{RC1} \ea \nit
in which $m_1, m_2$ and $m_3$ take on fixed values.
Thus in (\ref{RC1}), the sum over $m$ denotes a single sum, over $m_6$
for
example. The Racah coefficient $W$ involves four triad of angular
momenta
\be \label{RC2} (j_1, j_6, j_5), \quad (j_2, j_4, j_6), \quad
(j_3, j_4, j_5), \quad (j_1, j_2, j_3). \quad \e \nit

We intend to pursue the analogy of results like (\ref{x3}) in a
somewhat loose or qualitative way. Thus we consider the square
root factors in (\ref{RC1}) as being absorbed into the Racah coefficient
$W$,
and ignore signs.

We begin by comparing (\ref{RC1}) and (\ref{x3}). We have already
mentioned
the view of $c_{ijk}$ as a set of Clebsches for $ad \otimes ad
\rightarrow ad$
in a basis of Cartesian rather than angular momentum type.
Now we regard the the numerical factor $\fract{1}{2}$ on the right side
of (\ref{x3}) as a Racah coefficient with all six arguments equal to
$ad$.

Likewise, (\ref{kk2}) suggests that the Racah coefficients with five
arguments
$ad$ and its fifth argument $R_2$, in the place corresponding to $j_3$
in
(\ref{RC1}), takes the value
$\ell_2$, whilst (\ref{kk3}) suggests that the Racah coefficients with
five arguments $ad$ and its sixth argument $R_2$ takes the value $\ell_2
D_2/D$.

We wish here to make the point that, if one were to define Racah
coefficients systematically for $\g \in \cl{E}$, then it would be
expected
that
they would display full uniformity across $\cl{E}$. We have indicated a
few
simple examples in justification of this.
Also
\be \label{RC3} W(ad \; ad \; ad \; ad,\;  R_2 \; R_3)=
\frac{(11+D+\Delta)}{2D\Delta}. \e \nit

\end{document}